\documentclass[preprint,aps,draft]{revtex4}

\begin{document}

\title{Quantum Fluctuations in the Inflationary
Universe}

\author{Sang Pyo Kim}\email{sangkim@kunsan.ac.kr}

\affiliation{Department of Physics, Kunsan National University,
Kunsan 573-701, Korea\footnote{Permanent address}}

\affiliation{Asia Pacific Center for Theoretical Physics, Pohang
790-784, Korea}

\date{\today}
\begin{abstract}
The recent three-year WMAP data selects large-field models with
certain power-law potentials and small-field models for all
power-law potentials as consistent inflation models. We study the
large-field and small-field inflation model with a quadratic and a
quartic potential within the framework of quantum field theory in an
expanding Friedmann-Robertson-Walker universe. We find that quantum
fluctuations in the small-field model lead to a significant
contribution to the effective potential and may have non-negligible
effects on the slow-roll parameters predicted by classical theory.
\end{abstract}

\maketitle

\section{Introduction}

The inflation paradigm assumes a period of accelerated expansion of
the universe driven by the inflaton, a homogeneous scalar field or
condensate of the scalar field. Inflation models predict that the
density of the universe is close to the critical value, the geometry
of the universe is flat, the power spectrum of the primordial
density perturbations is nearly Harrison-Zeldovich scale-invariant
and the CMB radiations are homogeneous and isotropic. Now the
three-year WMAP data is able to select certain types of inflation
models and at the same time exclude many other models
\cite{spergel}.

Inflation models are classified into the single-field models and
multi-field models depending on the number of fields involved. The
single-field models have no internal space and are described by a
single scalar field and are further classified into the large-field
and small-field models \cite{dodelson}. The chaotic inflation
belongs to the former while the new inflation model belongs to the
latter. In large-field models the inflaton obtains the potential
energy from Planck scale quantum fluctuations, whereas in
small-field models it starts from a local maximum resulted from a
symmetry breaking. The WMAP data together with other astronomical
observations seems to favor the large-field and small-field models
\cite{spergel,martin,kinney,devega}.

In large-field or small-field models the inflaton can have the
energy scale requiring quantum theory for a proper framework. The
necessity of quantum theory for inflation dynamics may be understood
even through a quantum-to-classical transition from quantum gravity
\cite{banks} or a coherent state being close to a classical field
\cite{kim96}. Then the inflaton shows a characteristic behavior that
a classical background field drives inflation and its quantum
fluctuations lead to density perturbations. Quantum effects on
inflation models have recently been studied in small-field models
\cite{holman} and in large-field models \cite{boyanovsky}.

The main purpose of this paper is to study the inflation models
favored by the three-year WMAP data within semiclassical gravity.
Either semiclassical gravity from quantum gravity based on the
Wheeler-DeWitt equation \cite{banks} or quantum field theory in a
classical background spacetime may be used. As we concern more about
quantum fluctuations of fields than gravity itself, we shall employ
quantum field theory in a classical spacetime. Since the inflaton
evolves out of equilibrium as the universe expands rapidly during
the inflation period, nonequilibrium quantum field theory should be
used. One of nonequilibrium formalisms is the closed-time path
integral introduced long time ago by Schwinger and Keldysh
\cite{schwinger}. In this paper we shall use a canonical formalism
that unifies the functional Schr\"{o}dinger equation \cite{fun sch}
with the Liouville-von Neumann equation for density operator
\cite{kim00}. In this canonical formalism the semiclassical
Friedmann equation can keep an analogous form of the classical
equation. We find that, in contrast with relatively small quantum
fluctuations in large-field models, quantum fluctuations grow
exponentially due to the spinodal instability during slow-rolling in
small-field models and may contribute non-negligible amounts to the
slow-roll parameters predicted by the classical theory. We calculate
the effective small-roll parameters including quantum fluctuations,
compare with classical values and finally discuss the physical
implications in cosmology.

\section{Semiclassical Gravity}

Inflation models assume the existence of an inflaton (a homogeneous
scalar field or a condensate of the scalar field) whose energy
density leads a period of accelerated expansion of the universe. The
spacetime of the universe is assumed to have the homogeneous and
isotropic Friedmann-Robertson-Walker metric
\begin{eqnarray}
ds^2 = - dt^2 + a^2(t) d \Omega_3^2,
\end{eqnarray}
where $d \Omega_3^2$ is the metric for three-dimensional space at
fixed $a(t)$. In classical gravity the inflation dynamics is
described by the Friedmann equation (time-time component of the
Einstein equations) (in units of $G = c = 1$)
\begin{eqnarray}
\Bigl(H = \frac{\dot{a}}{a} \Bigr)^2 + \frac{k}{a^2} = \frac{8
\pi}{3 a^3} \Bigl(\dot{\phi}^2 + V(\phi) \Bigr),
\end{eqnarray}
and the inflaton field equation
\begin{eqnarray}
\ddot{\phi} + 3H \dot{\phi} + V'(\phi) = 0.
\end{eqnarray}
Here and hereafter overdots denote derivatives with respect to the
comoving time $t$.

The single-field inflation models compatible with WMAP data are
classified into large-field models in which the change of the
inflaton field during $N$-fold is $\Delta \phi \gg M_P$ ($M_P =
1/\sqrt{8 \pi G}$ being the Planck mass)  and small-field models in
which the change is $\Delta \leq M_P$ \cite{dodelson}. The chaotic
inflation with monomial potentials $V = M^4 (\phi/M_P)^p$ belongs to
the former class and the new inflation with potentials $V = M^4 [1 -
(\phi/\mu)^p]$ belongs to the latter class.  Here models are
characterized by the energy scale $M$ and power-law exponent $p$
and/or $\mu$. It is remarkable that the three-year WMAP data now
excludes $p > 3.1$ at 95 \% CL in the large-field models but allows
all the value of $p$ in the small-field models \cite{martin}. In the
large-field models the best fit with three-year data is $m^2 \phi^2$
but $\lambda \phi^4$ is not completely ruled out and still on the
edge of 68 \% CL.

Though most inflation models assume classical gravity as the
underlying theory, the proper theory should be semiclassical
gravity, in which the background spacetime is still classical but
matter fields are quantum. It is widely accepted that quantum
fluctuations are the seeds for structure formation, primordial black
holes and defect formation. Primordial density perturbations are
remnants of quantum fluctuations and quantum fluctuations may have
some imprints on CMB data such as the non-Gaussianity or on cosmic
gravitational waves. Furthermore, the energy scale of the inflaton
at the onset of the inflation belongs to a quantum regime. In the
new inflation models the inflaton slowly rolls down the potential
from an initial false vacuum toward the true vacuum of global
minimum whereas in the chaotic inflation models the infaton gains a
Planck-scale potential energy from quantum fluctuations. It is thus
legitimate to use semiclassical gravity or quantum field theory in
curved spacetimes to study the inflation dynamics. Further it would
be interesting to see how the inflation models work in semiclassical
gravity.

The very early stage of evolution of the universe would have been
described by quantum gravity, symbolically denoted by $\hat{G}_{\mu
\nu} = 8 \pi \hat{T}_{\mu \nu}$. It is expected to have a
quantum-to-classical transition of gravity: $\hat{G}_{\mu \nu} = 8
\pi \hat{T}_{\mu \nu} \Longrightarrow G_{\mu \nu} = 8 \pi \langle
\hat{T}_{\mu \nu} \rangle \Longrightarrow G_{\mu \nu} = 8 \pi T_{\mu
\nu}$ \cite{banks}. (From now on we drop the overhats for operators
unless they cause confusion.) The universe enters the semiclassical
gravity regime when the Planck scale gravity first decoheres and
becomes classical but matter fields still keep quantum nature. This
is, in the semiclassical gravity regime, the spacetime evolution is
governed by the Friedmann equation (in units of $G = c = \hbar = 1$)
\begin{eqnarray}
H^2 + \frac{k}{a^2} = \frac{8 \pi}{3a^3} \langle H_{\phi} \rangle.
\end{eqnarray}
Here the inflaton with the Hamiltonian,
\begin{eqnarray}
H_{\phi} = \int d^3 x  \Bigl[ \frac{\pi^2_{\phi}}{2a^3} +
\frac{a}{2} (\nabla \phi)^2 + a^3 V(\phi) \Bigr], \label{ham}
\end{eqnarray}
obeys the functional Schr\"{o}dinger equation \cite{fun sch}
\begin{eqnarray}
i \frac{\delta \Psi(\phi)}{\delta t} = H_{\phi} \Psi (\phi).
\end{eqnarray}

It should be noted that the inflaton as well as other matter fields
evolves out of equilibrium because the Hamiltonian (\ref{ham})
depends on the scale factor $a(t)$ of the universe, in particular,
when the universe undergoes an accelerated expansion phase. One may
use a criterion that a system evolves out of equilibrium when the
operator $\rho_H (t) = e^{- \beta H_{\phi}(t)}/Z_H$ deviates by a
large amount from the true density operator $\rho_I (t)$ in the
sense of $|| \rho_H (t) - \rho_I (t)||/||\rho_I (t)|| \gg 1$ with
respect to an appropriate measure $|| \cdot ||$ \cite{kim00}. On the
other hand, when $|| \rho_H (t) - \rho_I (t)||/||\rho_I (t)|| \leq
{\cal O} (1)$, the system evolves quasi-equilibrium. The true
density operator satisfies the Liouville -von Neumann equation
\begin{eqnarray}
i \frac{\delta \rho_I (t)}{\delta t}  + [\rho_I (t), H_{\phi} (t)] =
0.
\end{eqnarray}
The closed-time path integral by Schwinger and Keldysh may be one
method to handle this nonequilibrium evolution \cite{schwinger}. The
time-dependent functional Schr\"{o}dinger equation may be another
method that makes use of all useful properties of quantum mechanics
\cite{fun sch}. It is observed that the Liouville-von Neumann
equation may be used to solve not only the density operator but also
the time-dependent functional Schr\"{o}dinger equation
\cite{kim96,kim00}. In this paper we shall use the latter approach.

\section{Nonperturbative Method for Nonequilibrium Quantum Fields}

To treat the nonequilibrium quantum field, we first divide the field
into a classical background field and quantum fluctuations, $\phi(t,
{\bf x}) = \phi_c (t) + \phi_f (t, {\bf x})$, or use the (squeezed
or thermal) coherent state representation, $\langle \phi \rangle_C =
\phi_c$ and $\langle (\phi - \phi_c)^2 \rangle_C = \langle
\hat{\phi}_f^2 \rangle_{V/T}$ \cite{kim96,kim00}. We then decompose
the field and momentum into Fourier modes as
\begin{eqnarray}
\phi_f (t, {\bf x}) = \int \frac{d^3k}{(2 \pi)^3} \phi_{\bf k} (t)
e^{i {\bf k} \cdot {\bf x}}, \quad \phi_{\bf k} (t, {\bf x}) = \int
d^3x \phi_f (t, {\bf x}) e^{-i {\bf k} \cdot {\bf x}}.
\end{eqnarray}
In the oscillator representation
\begin{eqnarray}
\phi_{\bf k} (t) = \varphi_{\bf k} (t) a_{\bf k} (t) +
\varphi^*_{\bf k} (t) a^{\dagger}_{\bf k} (t), \quad \pi_{\bf k} (t)
= a^3 \Bigl( \dot{\varphi}_{\bf k} (t) a_{\bf k} (t) +
\dot{\varphi}^*_{\bf k} (t) a^{\dagger}_{\bf k} (t) \Bigr),
\label{con pair}
\end{eqnarray}
the Hamiltonian separates into a quadratic part $H_0$ and the part
$H_P$ higher than quadratic in $a_{\bf k}$ and  $a^{\dagger}_{\bf
k}$ as
\begin{eqnarray}
H_{\phi} (t, a_{\bf k}, a^{\dagger}_{\bf k}) = H_0(t, a_{\bf k},
a^{\dagger}_{\bf k}) + \lambda H_P (t, a_{\bf k}, a^{\dagger}_{\bf
k}).
\end{eqnarray}
It may be possible to find exactly the Green function for $H_0$,
\begin{eqnarray}
\Bigl[i \frac{\partial}{\partial t} - H_0 \Bigr] G_0 (t, {\bf x};
t', {\bf x}') = \delta(t-t') \delta({\bf x} - {\bf x}'),
\end{eqnarray}
in terms of an auxiliary field variable that obeys a mean-field type
equation, as will be shown in the next section. Then the full wave
functional is expanded perturbatively as
\begin{eqnarray}
\Psi(t, {\bf x}) = \Psi_0 (t, {\bf x}) + \lambda \int dt' d^3x' G_0
(t, {\bf x}; t', {\bf x}') H_P(t', {\bf x}') \Psi (t', {\bf x}').
\end{eqnarray}

We compare our formalism with the Hartree method, another canonical
formalism, which has been used to study large-field and small-field
inflation models in semiclassical gravity \cite{holman,boyanovsky}.
The Hartree method approximates the nonlinear Heisenberg equation by
a linear equation by factorizing nonlinear terms by powers of the
two-point correlation function, which includes some part of
higher-order loop corrections from the factorization. However, it
has a disadvantage that the other higher-order loop corrections
cannot be found beyond the linear evolution equation.

\section{Loop Corrections of Quantum Fluctuations}

In semiclassical gravity, the energy density and the effective
potential are obtained by taking the expectation value of the
corresponding operators with respect to the vacuum or thermal state
that satisfies the time-dependent functional Schr\"{o}dinger
equation:
\begin{eqnarray}
\rho_e (\phi) &=& \langle H_{\phi} (\phi_c(t) + \phi_f (t, {\bf
x})) \rangle_{V/T} = \rho_c (\phi_c) + \rho_q (\phi_c, \phi_f), \\
V_e (\phi) &=& \langle V (\phi_c(t) + \phi_f (t, {\bf x}))
\rangle_{V/T} = V_c (\phi_c) + V_q (\phi_c, \phi_f).
\end{eqnarray}
Here $\rho_c$, $V_c$ and $\rho_q$, $V_q$ denote the classical energy
density, potential and the loop corrections to the energy density
and potential, respectively. The expectation value with respect to,
for instance, a Gaussian vacuum state or a coherent-thermal state,
takes the simple form \cite{kim96,kim00}
\begin{eqnarray}
\langle \Bigl( \phi_c(t) + \phi_f (t, {\bf x}) \Bigr)^{2n}
\rangle_{V/T} = \sum_{k = 0}^{n} \frac{(2n)!}{2^k k! (2n - 2k)!}
\Bigl(\langle \phi_f^2 \rangle_{V/T} \Bigr)^k \phi_c^{2n - 2k},
\end{eqnarray}
where the two-point correlation function for quantum fluctuations is
given by
\begin{equation}
\langle \phi_f^2 \rangle_{V/T} = \int \frac{d^3{\bf k}}{(2\pi)^3}
(2n_{\bf k} (t_0) + 1) \varphi_{\bf k}^* (t) \varphi_{\bf k} (t).
\label{2 point}
\end{equation}
Here $n_{\bf k} (t_0)$ is the Bose-Einstein distribution of
fluctuations at the initial time $t_0$ and $\varphi_{\bf k} (t)$ is
an auxiliary field that will be determined below by the functional
Schr\"{o}dinger equation and/or Liouville-von Neumann equation. In
fact, the two-point correlation function is the evolution of that of
the initial time $t_0$ in the thermal or vacuum state.

In most of inflation models it is assumed that the classical field
$\phi_c$, which plays an order parameter here, dominates the energy
density and the potential and therefore determines the slow-roll
parameters for inflation. So the loop corrections of quantum
fluctuations to the slow-roll parameters may be written as
\begin{eqnarray}
\epsilon_e &=& \frac{M_p^2}{2} \Bigl(\frac{dV_e /d \phi_c}{V_e}
\Bigr)^2 = \epsilon_c \Bigl(\frac{1+ V_q'/V_c'}{1+ V_q/V_c} \Bigr)^2
= \epsilon_c(1 + \epsilon_q), \\
 \eta_e &=& M_p^2
\Bigl( \frac{d^2V_e /d \phi_c^2}{V_e} \Bigr) = \eta_c
\Bigl(\frac{1+V_q''/V_c''}{1+ V_q/V_c} \Bigr) = \eta_c(1 + \eta_q).
\end{eqnarray}
Note that $\epsilon_q$ and $\eta_q$ measure the relative amount of
quantum corrections to the slow-roll parameters. If $\epsilon_q$ and
$\eta_q$ are order of unity or larger than one, then quantum
fluctuations need to be included in the data analysis or dominate
over the classical parameters so the classical theory may not be
valid. In the latter case one needs a full nonperturbative quantum
theory such as the canonical method in Sec. 3 for inflation models.

In this paper we shall confine our attention to the potential of the
form $V = V_0 \pm m^2 \phi^2/2 + \lambda \phi^4/4!$, in which $V_0 =
M^4$, $- m^2 = - M^4/\mu^2$ and $\lambda = M_{\lambda}^4/\mu^4$ for
a small-field model while $V_0 =0$, $+m^2 = M^4/M_p^2$ and $\lambda
= M_{\lambda}^4/M_p^4$ for a large-field model, both of which are
favored by three-year WMAP data. The quartic term $\lambda
\phi^4/4!$ is necessary to stop the inflaton from rolling down
eternally in the small-field model but may be a minor modification
of the massive scalar field model in the large-field model. The
large-field and small-field model based on these potentials have a
sufficient $e$-folding necessary for successful inflation
\cite{koh}. The classical background field obeys the field equation
\begin{eqnarray}
\ddot{\phi}_c + 3H \dot{\phi}_c   \pm m^2 \phi_c + \frac{\lambda}{6}
\Bigl(\phi_c^2 + 3 \langle \phi_f^2 \rangle_{V/T} \Bigr) \phi_c = 0.
\label{cl eq}
\end{eqnarray}
In the Schr\"{o}dinger picture of Sec. 3, the auxiliary field obeys
the $c$-number equation \cite{kim00}
\begin{eqnarray}
\ddot{\varphi}_{\bf k} + 3H \dot{\varphi}_{\bf k} +
 \Bigl( \pm m^2 + \frac{{\bf k}^2}{a^2} + \frac{\lambda}{2} \phi_c^2
+ \frac{\lambda}{2} \langle \phi_f^2 \rangle_{V/T} \Bigr)
\varphi_{\bf k} = 0. \label{aux eq}
\end{eqnarray}
From the equal-time commutation relation $[\phi_{\bf k} (t),
\pi_{\bf k} (t)] = i$ in Eq. (\ref{con pair}), the complex solution
should satisfy the Wronskian condition
\begin{eqnarray}
a^3 (t) (\dot{\varphi}^*_{\bf k}(t) \varphi_{\bf k} (t) -
\dot{\varphi}_{\bf k} (t) \varphi_{\bf k}^* (t) ) = i. \label{wr
con}
\end{eqnarray}
The two-point correlation function is then obtained by putting the
complex solution $\varphi_{\bf k}$ in Eq. (\ref{aux eq}) satisfying
Eq. (\ref{wr con}) into Eq. (\ref{2 point}).

Finally, the loop corrections to the slow-roll parameters are given
by
\begin{eqnarray}
\frac{V_q}{V_c} &=& \frac{\frac{\lambda}{4} \langle \phi_f^2
\rangle_{V/T} \phi_c^2 \pm \frac{m^2}{2} \langle \phi_f^2
\rangle_{V/T} + \frac{\lambda}{8} (\langle \phi_f^2
\rangle_{V/T})^2}{V_0 \pm \frac{m^2}{2} \phi_c^2 +
\frac{\lambda}{24}\phi_c^4}, \\
\frac{V_q'}{V_c'} &=& \frac{\frac{\lambda}{2} \langle \phi_f^2
\rangle_{V/T} \phi_c}{\pm m^2 \phi_c +
\frac{\lambda}{6}\phi_c^3}, \\
\frac{V_q''}{V_c''} &=& \frac{\frac{\lambda}{2} \langle \phi_f^2
\rangle_{V/T}}{\pm m^2 + \frac{\lambda}{2}\phi_c^2}.
\end{eqnarray}
In the large-field model, the classical background field slowly
rolls down from a large initial value $\phi_c (t_0) \gg M_p$ as
$\phi_c (t) \approx \phi_c (t_0) e^{- \int m^2/3H}$ but quantum
fluctuations oscillate due to the positive frequency squared
$\omega_{\bf k}^2 (t) = m^2 + {\bf k}^2/a^2 + \lambda (\phi_c^2/2 +
 \langle \phi_f^2 \rangle_{V/T})/2$ as
\begin{eqnarray}
\varphi_{\bf k} (t) &\approx& \frac{1}{\sqrt{2 a^3 (t) \Omega_{\bf
k}
(t)}} e^{- i \int \Omega_{\bf k} (t)}, \nonumber\\
\Omega_{\bf k} (t) &=& \Biggl[\omega_{\bf k}^2 -
\Bigl(\frac{9}{4}H^2 + \frac{3}{2} \dot{H} \Bigr) + \Bigl(
\frac{1}{4} \Bigl(\frac{\dot{\Omega}_{\bf k}}{\Omega_{\bf k}}
\Bigr)^2 - \frac{1}{2} \Bigl( \frac{\dot{\Omega}_{\bf
k}}{\Omega_{\bf k}}\Bigr)^{\cdot} \Bigr) \Biggr]^{1/2}.
\end{eqnarray}
In the lowest-order WKB approximation $\Omega_{\bf k} \approx
\omega_{\bf k}$. Hence $\varphi^*_{\bf k} \varphi_{\bf k} \approx 1/
(2 a^3 \Omega_{\bf k})$ and quantum corrections during the inflation
period are suppressed as $V_q/V_c \approx V_q'/V_c' \approx
V_q''/V_c'' \approx \lambda \langle \phi_f^2 \rangle_{V/T}/(2m^2)
\ll 1$. It is shown that the loop corrections indeed change one
percent of classical slow-roll parameters \cite{boyanovsky}.

On the other hand, in the small-field model quantum fluctuations
grow exponentially due to the spinodal instability from the negative
curvature dominated by $-m^2$ at the onset of the inflation. This
growth of quantum fluctuations due to the spinodal instability
overcomes the damping due to the expansion of the universe.  At the
onset of inflation the inflaton slowly rolls down the potential, so
as long as the self-interaction is small compared with the negative
mass squared, the inflaton approximately becomes a scalar field with
the negative mass squared in an expanding spacetime. The scalar
field in the de Sitter spacetime $a(t) = e^{Ht}$ may shed light on
the dynamics of the inflaton. The mode equation
\begin{eqnarray}
\ddot{\varphi}_{\bf k} + 3H \dot{\varphi}_{\bf k} + \Bigl( {\bf k}^2
e^{- 2 Ht} - m^2 \Bigr) \varphi_{\bf k} = 0,
\end{eqnarray}
has the solution
\begin{eqnarray}
\varphi_{\bf k} = \sqrt{\frac{\pi}{4H}} e^{- 3Ht/2} H^{(1)}_{\nu} (k
e^{-Ht}/H), \quad \nu = \sqrt{\frac{9}{4} + \frac{m^2}{H^2}}.
\end{eqnarray}
At $t = - \infty$, all the modes are inside the Hubble horizon and
behave as the Minkowskian modes $\varphi_{\bf k} \approx e^{-i k
t}/\sqrt{2k}$. However, at later times, while ultraviolet modes,
which matters renormalization, are still inside the Hubble horizon,
infrared modes cross the Hubble horizon and grow as $|\varphi_{\bf
k}|^2 \approx \Gamma(\nu)(2H/k)^{2 \nu} e^{2 (\nu - 3/2)Ht}/(4 \pi
H)$. Therefore the loop corrections may not be negligible in
small-field models.

\section{Conclusion}

Cosmology now has become a science of precision in the sense that
measurments of CMB select the inflation models. In particular, the
three-year WMAP data favors large-field models with power-law
potentials for $p \leq 3.1$ and small-field models for all values of
$p$ \cite{martin,kinney}. This analysis is based on the classical
dynamics of the inflaton. However, at the onset of inflation, the
inflaton would be in a quantum state and afterward evolve out of
equilibrium due to a rapid expansion of the universe. It is likely
that quantum fluctuations modify the slow-roll parameters.

In this paper we have shown that the nonequilibrium quantum
evolution of the inflaton changes the slow-roll parameters as
$\epsilon_{e} = \epsilon_c(1 + \epsilon_q)$, $ \eta_{e} = \eta_c(1 +
\eta_q)$, etc. As the contribution of quantum fluctuations is
comparable to the classical one particulary for small-field models
and thus invalidates the classical inflation models, it would be
worthy to study inflation models in semiclassical gravity. It is
likely that the nonequilibrium quantum evolution may distinguish the
small-field models owing to the spinodal instability of quantum
fluctuations from the large-field models with small fluctuations. In
small-field models quantum fluctuations require a full
nonperturbative treatment. The detailed calculation of quantum
fluctuations on slow-roll parameters in the small-field and/or
large-field inflation models and the comparison with WMAP data and
other astronomical observations will be addressed in a future
publication.

\acknowledgements

The author would like to express his appreciation of warm
hospitality during CosPA 2006 at National Taiwan University. This
work was supported by the Korea Science and Engineering Foundation
under Grant No. R01-2005-000-10404-0.

\end{document}